\documentclass[preprint,showpacs,preprintnumbers,amsmath,amssymb]{revtex4}
\usepackage[dvips]{graphics}

\begin{document}
\draft
\title {Wigner function non-classicality as indicator of quantum chaos}
\author {A. Kowalewska-Kud{\l}aszyk}
\email{annakow@amu.edu.pl}
\affiliation{Nonlinear Optics Division, Institute of Physics, Adam Mickiewicz
 University, Umultowska 85, 61-614 Pozna\'n, Poland}
\author {J.K. Kalaga}
\affiliation{Nonlinear Optics Division, Institute of Physics, Adam Mickiewicz
 University, Umultowska 85, 61-614 Pozna\'n, Poland}
\date{\today}
\author {W. Leo\'nski}
\email{wleonski@amu.edu.pl}
\affiliation{Nonlinear Optics Division, Institute of Physics, Adam Mickiewicz
 University, Umultowska 85, 61-614 Pozna\'n, Poland}
\date{\today}

\begin{abstract}
We propose a Wigner function based parameter that can be used as an indicator of quantum chaos. This parameter is defined as "entropy" from the time-dependence of "non-classicallity" proposed in \cite{KZ04}. We perform our considerations for the system of damped nonlinear (Kerr-like) oscillator excited by a series of ultra-short external pulses.
\end{abstract}

\maketitle

\section{Introduction}
The systems demonstrating chaotic behaviour in their dynamics are still the subject of much attention. Especially, there has been a great deal 
of interest in exploring classical dynamics of chaotic systems. 
For such systems certain methods allowing distinction between the regions of regular dynamics from those of chaotic one are well developed and widely used.
It is still of special importance to find the adequate methods for analysis of the systems whose quantum dynamics exhibits
chaotic behaviour. It is of interest not only from the cognitive point of view but is also for the development of quantum information theory methods. It is known that chaotic behaviour in the dynamics of a quantum system should destroy the entanglement (which is the essential point in the quantum computing) between the quantum states, but surprisingly in some cases quantum chaos can even enhance the stability of quantum computation \cite{PZ01}.

Many attempts to address the problem of quantum chaos have already been made.
For example, it is known that there is a correspondence between the statistics of eigenvalues and eigenvectors of quantized classically chaotic systems  and the canonical ensembles of the random matrix theory \cite{HMABH07,SR01,BGS84,BT77,I87,KMH88,HZ90}. The distances between successive eigenvalues of quantized chaotic system have the same probability distribution as the successive distances between the eigenvalues of random matrix.
There is also a method based on the fidelity decay, indicating the existence of chaos in dynamics of quantum systems \cite{P84,EWLC02,WLT02}. 
The time-evolution of the fidelity  between the state evolving under the unitary mapping procedure and the same initial state evolving under the same map but subjected to some additional tiny perturbations, in the chaotic region exhibits an exponential decay \cite{WLT02}. It means that there is a significant difference between these two quantum states and that the dynamics of the quantum system is sensitive to the initial conditions. This sensitivity is a characteristic feature of the chaotic dynamics.

In the present article we would like to apply the quantum parameter based on the Wigner function (and connected to its negative volume), introduced in \cite{KZ04} and referred to as the ''non-classicallity parameter'', to the problem of a quantum counterpart of the classically chaotic system dynamics. We introduce an entropic measure based on this parameter and show it to be sensitive to changes in the regions where the classical counterpart of the system described is chaotic. These changes indicate that the final state of the quantum system in some regions is also very sensitive to the initial condition variations and hence, very tiny perturbation of the initial conditions lead to considerable changes in the final quantum state.
To investigate such phenomena we would apply the recurrence plots analysis. It will allow us to confirm our statements about the chaotic behaviour (in the classical sense) of the strictly quantum parameter defined from the Wigner function of the system described.

\section{The model}
We consider a system composed of an anharmonic oscillator driven by a series of ultra-short external pulses. Various aspects of such a system's dynamics have been discussed in many papers (for example see \cite{LT94,LDT95,Z81} and the referrences quoted therein). It is known that depending on the values of the parameters used, the system can exhibit regular or chaotic dynamics \cite{L96}. 
The problem of the classical dynamics and that of the quantum counterpart of this system is also discussed in that paper. 

The system is supposed to be initially in the vacuum state. In the interaction picture, the time evolution of the system under consideration is governed by the following Hamiltonian (we use units $\hbar=1$):
\begin{equation}
\hat{H}=\hat{H}_{NL}+\hat{H}_{K}\,\,,
\end{equation}
where $\hat{H}_{NL}$ describes the evolution between subsequent pulses and $\hat{H}_K$ -- the evolution initiated by an ultra-short pulse:
\begin{eqnarray}
\hat{H}_{NL}&=&\frac{\chi}{2}\left(\hat{a}^+\right)^2\hat{a}^2\,\,,\\
\hat{H}_{K}&=&\epsilon\left(\hat{a}^{+}+\hat{a}\right)\sum\limits_{k=1}^\infty\delta(t-kT)\,\,.
\label{h3}
\end{eqnarray}
Operators $\hat{a}^+$ and $\hat{a}$ are those of photon creation and annihilation, respectively; $\chi$ describes the non-linearity of the oscillator (for Kerr medium it is the third order susceptibility) and in further considerations it is set to $1$; $\epsilon$ is the strength of the external pulse -- nonlinear oscillator interaction, and $T$ is the time between two subsequent pulses. Assuming that time $T$ exceeds significantly the the duration of single pulse, we can model the series of these ultra-short pulses by a series of Dirac-delta functions.

In the present considerations we deal with the damping case only. It seems to be a more realistic situation to let an oscillator to interact with the enivronment
than simply deal with a no-damping case. In real physical situations, the influence of the environment (represented by the damping process) may significantly change the system's dynamics. For the system considered the influence of the damping process makes the system's dynamics regular for a wider range of external excitation strength than in the no-damping case. As the system is damped in the chaotic region fewer states are involved in the system's dynamics, which significantly simplifies numerical calculations. When we do not include the damping process, in the chaotic region many more states have to be considered and additionally, the chaotic behaviour in the quantum system considered is visible for longer times. If we look closer at the numerical complexity of the problem it becomes clear that for the cases without damping it would be more suitable to analyse for example a fidelity decay as a signture of chaotic dynamics in a quantum system rather than to deal with a Wigner function based parameter. When dealing with a Wigner-based parameter (which has to be calculated and analysed after each external kick) it would be more desirable to work with smaller basis and to have an opportunity to observe chaotic behaviour of the system in shorter times as well.

For the system considered the density matrix approach is necessary and we use the formalism proposed in \cite{MH86} and used for example in \cite{L96}.
The time evolution of the density matrix is governed by the following master equation:
\begin{equation}
\frac{d\hat{\rho}}{d t}=-i\frac{\chi}{2}\left[(\hat{a}^{+})^2\hat{a}^2,\hat{\rho}\right]+\frac{\gamma}{2}\left(2\hat{a}\hat{\rho}\hat{a}^{+}-\hat{a}^{+}\hat{a}\hat{\rho}-\hat{\rho}\hat{a}^{+}\hat{a}\right)\,\,\,,
\label{master}
\end{equation}
that can be solved analytically (for example see \cite{MH86}) or numerically.
In fact, equation (\ref{master}) describes the free evolution of the damped system between two subsequent pulses. 
To include the interaction with the external field, one should apply to the density matrix (after its free evolution during the time T) the operator $\hat{U}_K$  of the form:
\begin{equation}
\hat{U}_K=e^{-i\epsilon(\hat{a}^{+}+\hat{a})}\,\,\, .
\label{Uk}
\end{equation}
Consequently, the whole dynamics will be described successively by the evolution (during the time T) according to the master equation (\ref{master}) and the ''kicked'' operator (\ref{Uk}). The initial state of the system considered is the ground state $\rho(t=0)=|0\rangle\langle 0|$.

It is known that the classical kicked Kerr oscillator, depending on the values of the parameters used, can exhibit regions with classically regular or chaotic behaviour. Appropriate bifurcation diagrams of the mean classical energy as a function of the external coupling parameter $\epsilon$, for weak and strong damping cases  have been presented in \cite{L96}. 
The procedure used for obtaining such diagrams first requires the explicit form of the equation of motion for the annihilation operator for the time between the subsequent external pulses. The solution of this equation can be expressed in a simple analytical form, because at first we neglect the damping process, and consequently, the number of photons $\hat{n}=\hat{a}^{+}\hat{a}$ is conserved during the evolution. Then, we include the influence of the external pulse (described by the action of the operator (\ref{Uk}) which in fact is a shift operator. The final point of the procedure is to replace all the operators ($\hat{a}^+$, $\hat{a}$) by the complex numbers ($\alpha^\star$, $\alpha$) and the damping rate $\gamma$ can be introduced at this point. Finally, the equation for $\alpha$ has the form:
\begin{equation}
\alpha_{k+1}=\left(\alpha_k-i\epsilon\right)e^{-i\left(\chi|\alpha_k-i\epsilon|^2-i\gamma\right)T}\,\,\, ,
\label{alpha}
\end{equation}
and the formula for the classical mean energy $|\alpha|^2$ needed for the construction of the bifurcation diagram can be obtained easily.
A classical state which would be a classical counterpart of the initial system state (the vacuum state) can be simulated as in \cite{MH86}. To introduce the initial condition we have created an ensemble of classical trajectories whose starting points had been randomly chosen from a circle of radius $0.5$ and centred at $\alpha=0$ and after that the average trajectory has been treated as a classical counterpart of a quantum one. 

In the present paper we shall concentrate on the case when damping constant $\gamma=0.1$. 
The bifurcation diagram for a classical mean energy (Fig.1) shows that for a wide range of the external kicks strength $\epsilon<\sim 0.9$, one can observe the regular dynamics of the classical oscillator. The chaotic behaviour appears when $\epsilon$ exceeds the value of $\sim 0.9$. Additionally, the bifurcation (starting from $\epsilon\approx 0.7$) appears in the diagram. This situation corresponds to that discussed in \cite{L96}. 
Moreover, one should remember that the regular and chaotic behaviour visible in the bifurcation diagram concerns the classical dynamics of the anharmonic oscillator.

In the present paper we will try to answer the question about the regularity or chaocity in the dynamics of the quantum system whose classical counterpart behaves regularly or chaotic. In other words we will try to make use of a quantum parameter (characterising the quantum system) to identify the regions of regular quantum dynamics and those of a quantum chaotic nature.
We have chosen a quantum parameter (reffered to as ''nonclassicality'') based on the negativity of the Wigner function \cite{KZ04} for the analysis of quantum dynamics of the kicked nonlinear oscillator. As claimed in \cite{KZ04}, this parameter is connected to the quantum character of the states analysed  and consequently, we shall check its usefulness as an indicator of quantum chaotic behaviour.

\section{The evolution of the non-classicallity indicator}
 
This paper is devoted to the applicability of the "non-classicallity" parameter as an indicator of quantum chaos.
This parameter has been introduced in \cite{KZ04} and is related to the negative volume of the Wigner function. 
It is known that the coherent states minimise the uncertainty principle and in this sense they are considered as classical states. When we talk about the ''non-classicallity'' of the states we understand this ''non-classicallity'' as the deviation from the coherent states.
It is known that the Wigner function is a quasi-probability function that represents a quantum state in the coherent state basis. The Wigner function is joined with the symmetrically ordered bosonic operators. For a state described by the density matrix, $\rho$ is defined as \cite{CG69}:
\begin{equation}
W(\alpha)=2Tr\left[\hat{D}^{-1}(\alpha)\hat{\rho}\hat{D}(\alpha)\hat{P}\right]\,\,\,,
\label{w1}
\end{equation}
where $\hat{D}(\alpha)=\exp[\alpha\hat{a}^{+}-\alpha^\star\hat{a}]$ is the displacement operator and $\hat{P}=\exp[i\pi\hat{a}^+\hat{a}]$ is the parity operator. The Wigner function can take both positive and negative values.  
The non-classicallity indicator proposed and discussed in \cite{KZ04} is defined as follows:

\begin{equation}
\delta(\rho)=\int\left[|W(\alpha)|-W(\alpha)\right]d\alpha\,\,\,.
\label{d1}
\end{equation}
The value of $\delta$ depends on the volume of the negative part of the Wigner function and is equal to zero for the coherent and squeezed vacuum states for which the Wigner function is non-negative \cite{H74}. In this sense, the $\delta(\rho)$ parameter describes how much the  quantum state considered differs from the coherent one and in consequence, form the ''classical'' state. It was shown by the authors of the $\delta(\rho)$ definition, that the value of this parameter for the Fock states $|n\rangle$ grows monotonically with $n$ and consequently, the higher the value of $n$ the greater the deviation of the state $|n\rangle$ from the coherent state. Moreover, the value of $\delta(\rho)$ does not depend on the degree of squeezing. 

We have analysed the time evolution of the parameter $\delta(\rho)$ defined as in (\ref{d1}), concentrating on the parameters describing our system ($T$, $\chi$, $\gamma$), used for analysis of the bifurcation diagram. From the whole range of external field -- nonlinear system coupling strengths $\epsilon$ we have chosen the values corresponding to the classically regular and chaotic regions.

Moreover, in this paper we use $\delta_{n}(\rho)=\delta(\rho)/n$ instead of $\delta(\rho)$, which means that the non-classicallity is divided by the mean number of photons after each external kick - so we use the parameter which can be more convenient in further considerations when the number of photons grows considerably (regions of chaos).
In the regions of regular dynamics it causes no significant changes in the values of $\delta(\rho)$. 
Figures 2 and 3  present the time-evolution of $\delta_n(\rho)$ for the excitation strengths corresponding to the classically regular $\epsilon=0.2; 0.3; 0.4$ -- Fig.2a and chaotic  $\epsilon=1.24$ --- Fig.2b dynamics. 
Thus, we have found that whenever the classical system exhibits regular dynamics, the changes in $\delta_n(\rho)$ are regular, and 
depending on the $\gamma$ value they may be slower or faster damped. The character of time-evolution does not change with increasing value of $\epsilon$. The regular damped oscillations (whose frequency depends on the value of $\epsilon$) occur -- this behaviour can be seen in Fig.2a, where we have plotted $\delta(\rho)$ versus time for various values of external kick strength $\epsilon$. 
One should keep in mind that for such values of $\epsilon$ the classical counterpart of our quantum system behaves regularly.
We have plotted the time dependence of the mean energy 
of the classical system in that regular region -- Fig.3a (for recollection -- an average of randomly choosen $10^5$ -- in our case -- trajectories starting from a circle of radius 0.5 centered at $Re(\alpha)=Im(\alpha)=0$). We can see that indeed, the mean energy of the system (found for a sufficiently long period of time) does not vary significantly with changes in the external pulses strength. For comparison, when we plot the mean number of photons of the quantum system --- Fig.3b --- (which would be a kind of analog of the mean energy of the classical system) we can observe a similar behaviour of this quantity. 
Small changes in the kicking strength cause the same small changes in the final mean number of photons $\langle\hat{n}\rangle$ in the system.
For the case of strong external pulses ($\epsilon>0.9$ -- chaotic region) the situation changes considerably.
The changes in the values of the system's final energy are more sensitive to the changes in the external pumping strength. While for the case of weak excitations a change in $\epsilon$ of $\approx 0.1$ causes a change in the mean number of photons of 
$\approx 0.09$ for $\epsilon > 0.9$ the same changes in the pump strength cause $10\times$ larger increase in the mean  number of photons generated in the process. Moreover, the energy of the quantum system initially increases rapidly (as in its classical counterpart) for the values of $\epsilon$ corresponding to the "deep chaos" case in the classical system.
We can clearly see the similarities of the essential features of the two quantities discussed here (mean energy and mean number of photons). The values of the energy for the classical and quantum system's are not exactly the same but the character of their changes is similar in both (classical and quantum) models.

More interesting is the situation when the pulses are sufficiently strong to lead to chaotic behaviour in the classical counterpart of the system described. We have chosen the kick strength $\epsilon=1.24$. It is seen from the bifurcation diagram that this value is high enough to put the classical system into a region of deep chaos. For comparison, we can analyse the behaviour of the quantum counterpart of the kicked oscillator for the same values of the other parameters as in the previously discussed case. 
Fig.2b presents the evolution of $\delta_n(\rho)$ characterising the quantum system for the parameters which caused the classical counterpart of the system analysed to behave chaoticly. It is seen that there are no regular oscillations any more. Irregularities apparent in the time-evolution of $\delta_n(\rho)$ indicate that the final state of the quantum system considered cannot be well defined and changes significantly from one external pulse to another. Many states with various photon numbers are involved in the process and probability none of them is dominant. We can see that the parameter $\delta_n(\rho)$ indicates almost instantaneous irregular changes. 
The range of values reached by $\delta_n(\rho)$ indicates that the Wigner functions have noticeable negative parts (which is a measure of the quantumnes of the state). The negative fractions of $W$ also change from one pulse to another.  

\section{Recurrence Plots Analysis}
We have identified irregular changes in the quantum parameter $\delta_n(\rho)$ for the cases corresponding to the situation when the classical counterpart of the system exhibits chaotic behaviour. We suppose that these irregularities are connected with the fact that the quantum system dynamics is also chaotic. To confirm our supposition we will analyse the time dependence of the "non-classicallity" parameter via the recurrence plots (RP) method, which can be easily applied for a short time series.

The idea of recurrence plots comes from \cite{EKR87} and was developed and widely used in nonlinear data analysis. This method can be used to solve problems related to medicine (see for example \cite{AJKLS06,BZB02}), economy \cite{HZU01,FFMM03}, geophysics \cite{CM03,CM07}, astronomy \cite{DT07,ZP06} and many other fields, whenever there is a need to analyse a time varying signal coming from a nonlinear system. The basic idea of RP is to graphically present the times at which the system analysed recurs, or more precisely, the times at which the system's trajectory presented in an appropriate phase space returns (with some approximation) to the same area. RP allows analysis of nonlinear signals from the systems whose trajectories generally exist in a many-dimensional phase space via a 2-dimensional plot. 

To construct RP we need to determine the binary matrix $R$. Its elements are defined as \cite{EKR87}:
\begin{equation}
R_{i,j} = \Theta(\epsilon_{thr} - ||\vec{x}_i - \vec{x}_j||)\hspace{1cm} i,j=1\cdots, N\hspace{1cm} \vec{x}_{i}\in\Re\,\, ,
\end{equation}
where $\Theta$ is the Heviside function, $\epsilon_{thr}$ is the threshold parameter and $||\cdot ||$ denotes the norm.
This norm allows determination of the distance between two points. The trajectory of the system (obatined as time series of some system's parameter --- for the case discussed here it is $\delta_n(\rho)$) is first of all reconstructed in an apropriate phase space (whose dimension is determined before) at time t. Consequently, $\vec{x}_i - \vec{x}_j$ measures the distance between the points of the so reconstructed trajectory.
If two points fall inside the same region (sphere) according to $\epsilon_{thr}$, we would label them by $1$, otherwise they are labelled as $0$.
In consequence, one obtains a square matrix with zeroes and ones, or graphically with black and white points. 

The matrix can be analysed using the {\em Recurrence Quantification Analysis} (RQA) introduced by Zbilut and Webber Jr \cite{ZW92,ZGW98}, based on examination of diagonal structures of RP. This method has been extended by Marwan \cite{MWMSK02},who has proposed new measures of complexity based on vertical structures in RP.
In general, long diagonal lines in RP are characteristic of periodical orbits, whereas homogeneously distributed black points indicate white noise. On the other hand, the system exhibiting chaotic dynamics will produce shorter diagonal structures in RP and isolated black points among them. Moreover, the vertical structures suggest the existence of laminar states (states which do not change or change very slowly with time \cite{MWMSK02}).

In this paper we have applied the RQA method for analysing the time varying "non-classicallity" parameter $\delta_n(\rho)$. We have concentrated on the regions where the classical counterpart of the system analysed behaves chaotically and we suppose that the quantum system exhibits chaotic dynamics as well. For the  damping constant $\gamma=0.1$ we have obtained irregular changes in $\delta_n(\rho)$ (see Fig.2b). RP would help us determine whether the system characterised by the time varying parameter $\delta_n(\rho)$ exhibits chaotic dynamics or not. 
To generate the RPs we have used a CRP toolbox by Marwan \cite{MRTK07}. 

The first task is to reconstruct the phase space trajectory of the system using a time series of a measured quantity (in our case the "non-classicallity" of the system).
It can be done using a time-delay method, which involves the use of the apropriately chosen minimal sufficient dimension for reconstructing the original system's trajectory.
At first, we have estimated the optimal time delay $\tau$ (using mutual information function) as $\tau=1$.
It has to be chosen in such a way that the linear dependencies between two subsequent vectors in a reconstructed phase space are reduced. It can be realized by finding the minimum of 
the mutual information function which describes the joint probability of finding the time series value at the $i$-th interval and after a time $\tau$ in the $j$-th interval \cite{FS86}. 

Moreover, we have found the embedding dimension as $d=10$. We have used for this purpose the false nearest neighbours function that counts the number of points in the nearest neighbourhood of a specified point. A point marked as a neighbour of another one in lower dimension may not belong to the neigbourhood of this point in a higher dimension phase space - such a point is called a false neighbour and the false nearest neighbours function allows choosing such a phase space dimension at which the number of false neighbours vanishes.

For these parameters we have generated the RP presented in Fig.4a. We can see a pattern that is characteristic of chaotic dynamics --- some diagonal lines are visible. Their length distribution is plotted in Fig.4b. The maximum length of the diagonal line is 26 points. Although the lines of shorter lengths are dominant, there is a significant amount of lines with lengths larger than 2 points. Formation of diagonal lines is characteristic of chaotic dynamics of the parameter analysed and their lengths are related to the value of the largest Lyapunov exponent of a chaotic system \cite{EKR87,TRKMAA02}.
From RQA we have obtained  the average diagonal line length of 4 points. 
It is worth stressing that most of the points that are present in the RP form diagonal structures -- the percentage of these points is  $\approx 92\%$. Therefore, we can say that the process of non-classical states generation initiated by external pulses (appearing as an increase in $\delta_n(\rho)$) during the time of the evolution, for the excitations sufficiently strong to cause chaos in the classical counterpart of the system considered, is rather of deterministic nature. Stochastic processes are characterised by isolated points or very short lines (mainly composed of 2 points) and a low value of DET - contrary to the case discussed here.\\
Therefore, we can conclude that the parameter $\delta_n(\rho)$ can be treated as an indicator of chaos generated in quantum systems.

\section{The entropic measure of changes in non-classicallity parameter.}

As we are interested in a strictly quantum parameter that would allow us to determine whether the system is in the regular or chaotic region of its dynamics, we would use the definition of the non-classicallity parameter $\delta_n(\rho)$ and propose a quantity (having the physical sense of entropy) that measures the changes in $\delta_n(\rho)$. We can define this entropy in a few steps:
\begin{enumerate}
\item We define the Fourier transform of the $\delta_n(\rho)$ as:
\begin{equation}
{\cal F}(\omega)=\sum\limits_{t_{min}}^{t_{max}}\delta_{n}(t)e^{-i\omega t} \,\,\,.
\label{eq2}
\end{equation}
\item Then we calculate a "power spectrum"  ${\cal P}=|{\cal F}(\omega)|^2$ and normalise ${\cal P}$ to get  ${\cal P}_N(\omega)$. 
\item Finally, we define the entropic measure of changes in the non-classicallity indicator through the power spectrum as:
\begin{equation}
E=-\sum_{\omega}{\cal P}_N(\omega)\ln\left({\cal P}_N(\omega)\right) \,\,\,.
\label{eq3}
\end{equation}  
\end{enumerate}

In consequence, we obtain one value of entropy for the specific coupling strength $\epsilon$. The entropy value equal to zero would indicate that the non-classicallity parameter $\delta_n(\rho)$ does not change in the whole range of time considered. The increase in the value of $E$ and its irregular changes  for various values of coupling strength would indicate  significant changes in the time dependence of the $\delta_n(\rho)$ parameter. 
From the calculations of the above defined entropy, we have found that for the regions of regular system dynamics the changes in the value of the entropy with $\epsilon$ are smooth in character (Fig.5) indicating regular oscillations of $\delta_n(\rho)$. Changes in the value of entropy arise from the fact that the frequency of $\delta_n(\rho)$ oscillations change with $\epsilon$ and for higher values of $\epsilon$ more oscillations of $\delta_n(\rho)$ would appear prior to their vanishing (see Fig.2a). Additionally, with higher values of $\epsilon$ more states are involved in the process. But as long as the changes in $E$ are smooth, the system's dynamics is regular. 
 When the system is in the deep chaotic region we see significant and irregular changes in the value of entropy. These irregularities (appearing for excitation strengths corresponding to the classical deep chaos) are visible in Fig.5 (from $\epsilon\approx 1$). Therefore, we can see that the entropy $E$ (that is a strictly quantum parameter) can indicate irregular changes in the dynamics of the quantum system described. These irregularities appear for the values of $\epsilon$ leading to the chaotic behaviour of the classical counterpart of the quantum system described. We have already analysed these regions via the recurrence plots of the time varying $\delta_n(\rho)$ parameter, and we have identified the regions corresponding to the classical deep chaos as being chaotic also for the quantum system.
Therefore, we can conclude that the regions of irregular changes in the entropic parameter proposed are those of chaotic nature, indicating the presence of quantum chaos phenomena.

\section{Conclusions}
The quantum systems, whose classical counterparts exhibit the chaotic behaviour are still widely explored. Their analysis is expected to answer the question whether the chaotic dynamics of a classical system would persist in the dynamics of its quantum counterpart. 
It is known that for the classical systems there are well known methods allowing distinction between the regular and chaotic regions, whereas for the quantum models such methods are still being developed. Some of them have been already proposed and discussed, and we have mentioned them in the introduction of this article. 

In this paper we have proposed an entropic parameter $E$ (of strictly quantum nature) based on the {\em non-classicallity} proposed in \cite{KZ04}, connected with the Wigner function. We have shown that this parameter used for the analysis of the quantum system which in the classical limit can exhibit chaotic dynamics, undergoes rapid and significant changes with coupling strength $\epsilon$, in the regions corresponding to classical chaos. As an example we have analysed  a quantum Kerr-like oscillator externally kicked by the ultra-short pulses. It is well known that nonlinear quantum systems can be used as the components of systems exhibiting quantum entanglement and as such can be considered as a tool in quantum informatics. It is therefore of special interest to explore simple methods allowing the identification of the regular or chaotic regions in dynamics of such nonlinear quantum systems.
The behaviour of the entropic measure E for such a system indicates that the final state of the quantum system in the regions corresponding to the deep classical chaos cannot be well established and changes irregularly from one pulse to another. The entropic parameter, apart from the sensitivity to tiny perturbations, reveals irregular dynamics even for short times.  
We have analysed this short-time dependence of the non-classicality parameter (which is the basis of the entropic measure) using the RP method -- a tool used in time-series analysis. We have concluded that the dynamics of changes in the non-classicality of the state generated in the process discussed for the regions corresponding to classical chaos can be treated as being chaotic as well.
Therefore, we believe that the parameter $E$ proposed in this paper could be a useful tool for investigation of systems exhibiting quantum chaotic phenomena.

\newpage
\pagestyle{empty}
\begin{figure}[p]

\resizebox{14cm}{8cm}
                {\includegraphics{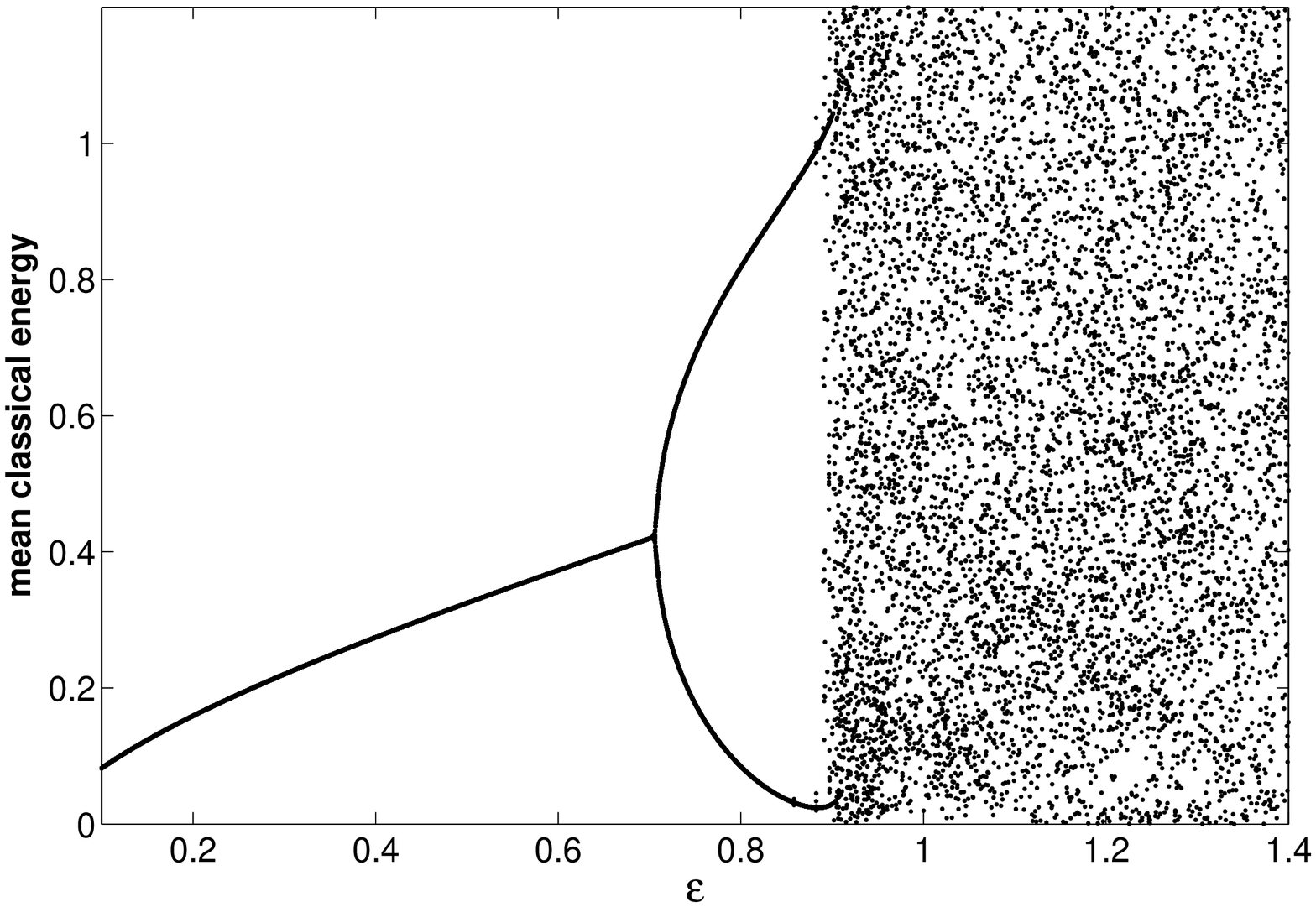}}
\caption{Bifurcation diagrams for $\gamma=0.1$.  The other parameters are: $\chi=1$, $T=\pi$.  }
\end{figure}

\newpage
\begin{figure}[p]
\resizebox{16cm}{10cm}
                {\includegraphics{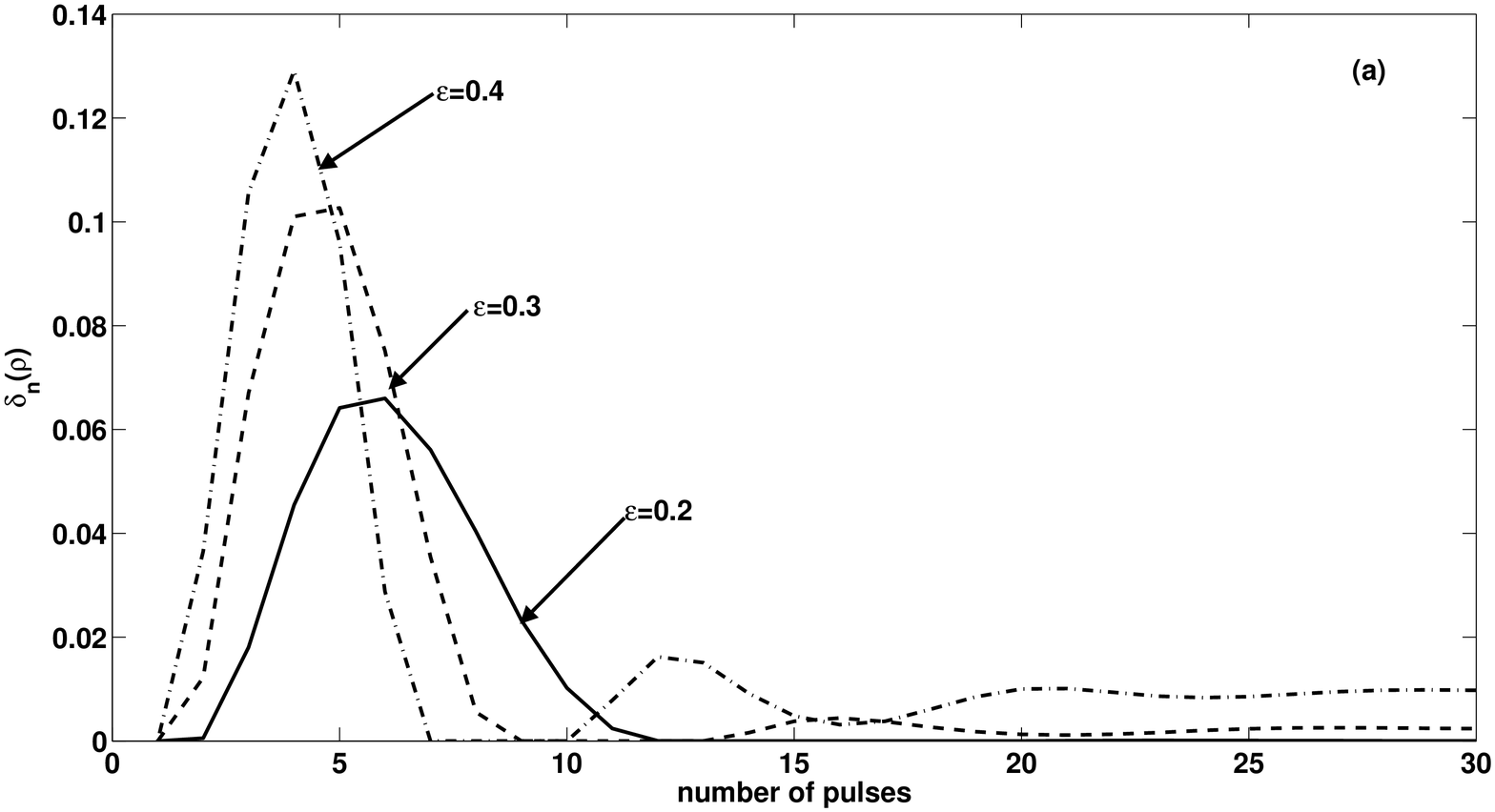}}
\resizebox{16cm}{10cm}
                {\includegraphics{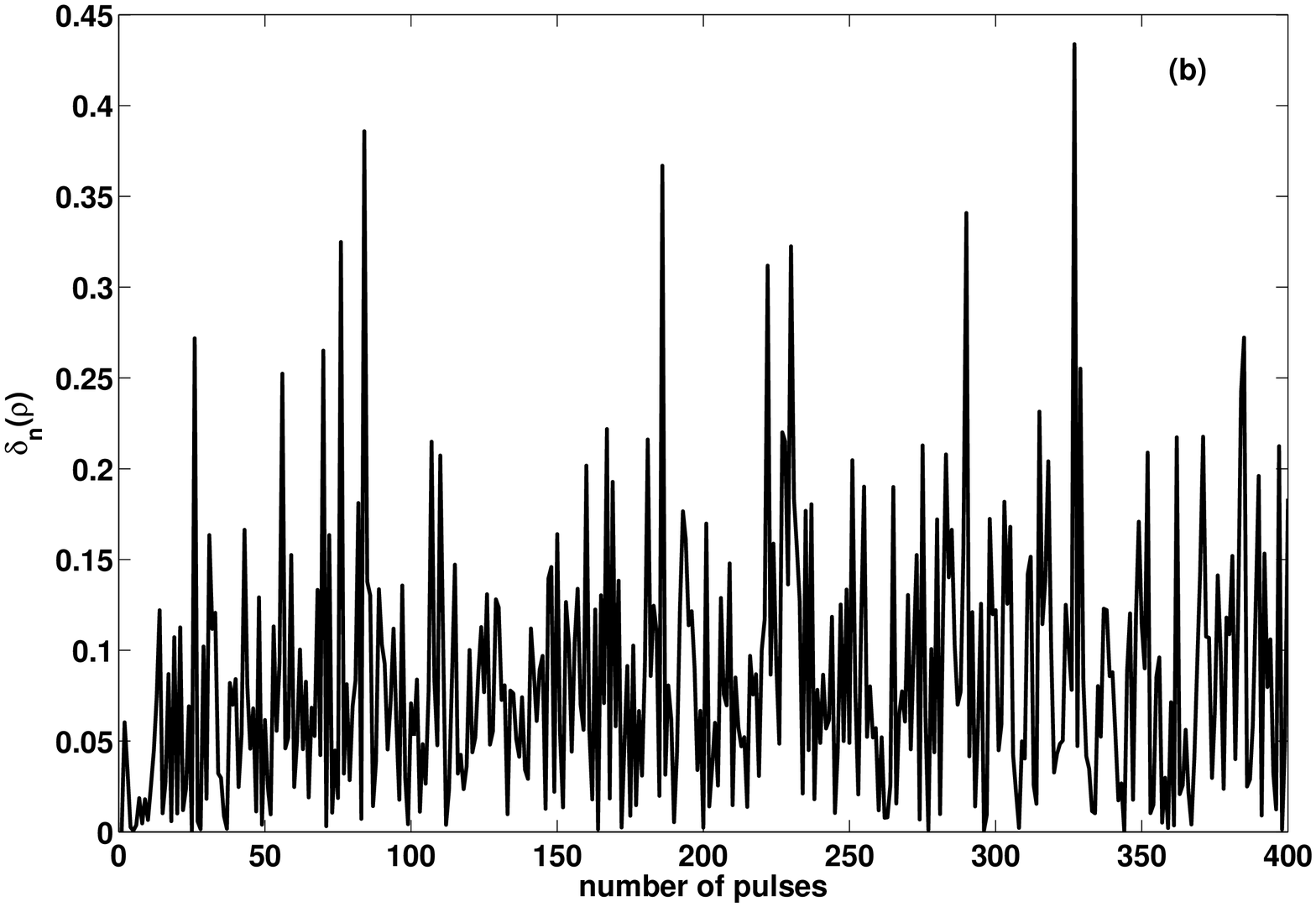}}
\caption{Time dependence of the ''non-classicallity'' parameter $\delta_n(\rho)$ for various values of $\epsilon$, at (a) $\epsilon=0.2;0.3;0.4$; at (b) $\epsilon=1.24$. Other parameters: $T=\pi$ and  $\gamma=0.1$} 
\end{figure}

\newpage
\begin{figure}[p]
\resizebox{16cm}{10cm}
                {\includegraphics{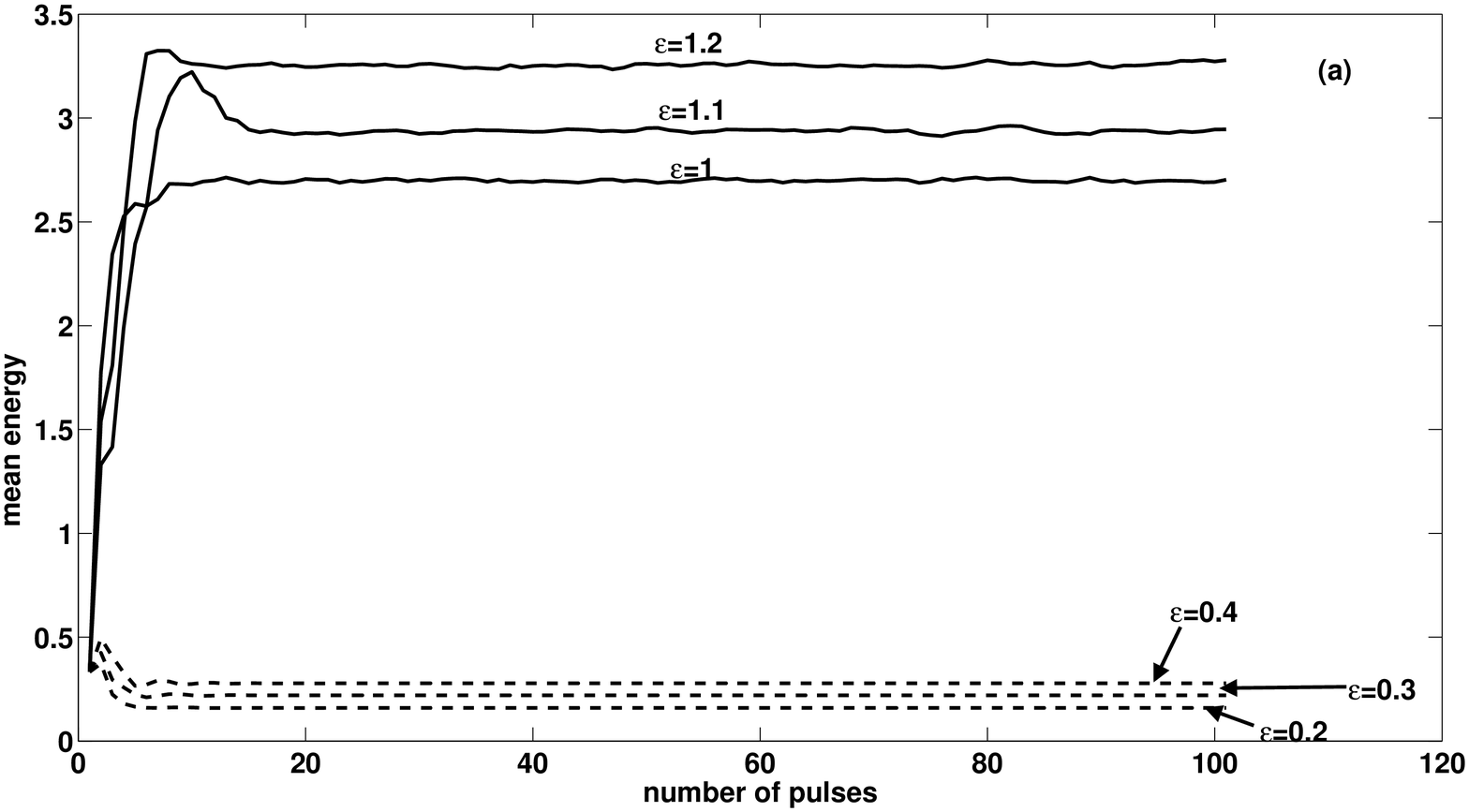}}
\resizebox{16cm}{10cm}
                {\includegraphics{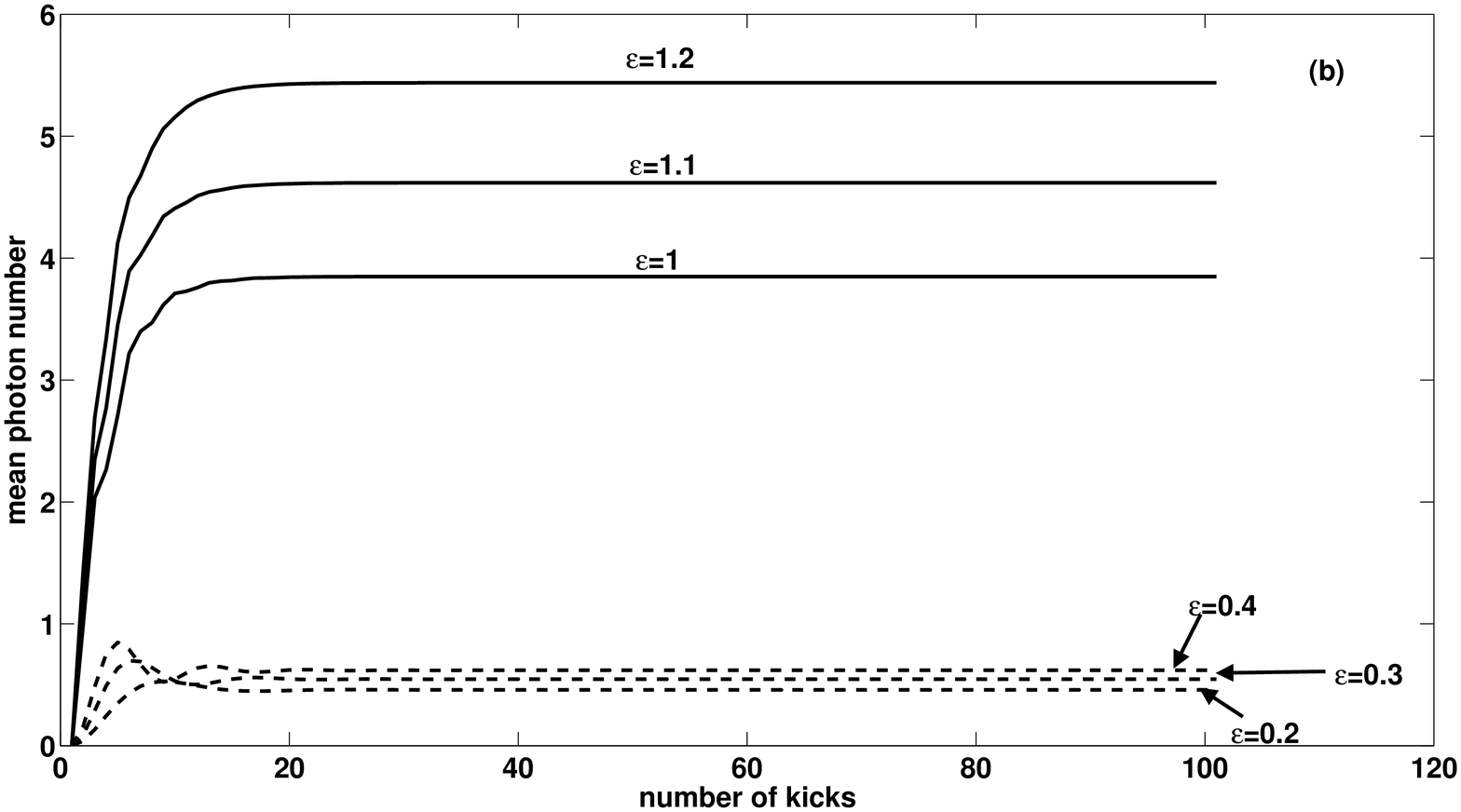}}
\caption{Time dependence of the mean energy of the classical system (a) and the mean number of photons (b) for various values of $\epsilon$. The parameters: $T$ and  $\gamma$ are the same as in Fig.2} 
\end{figure}

\newpage
\begin{figure}
\resizebox{16cm}{10cm}
                {\includegraphics{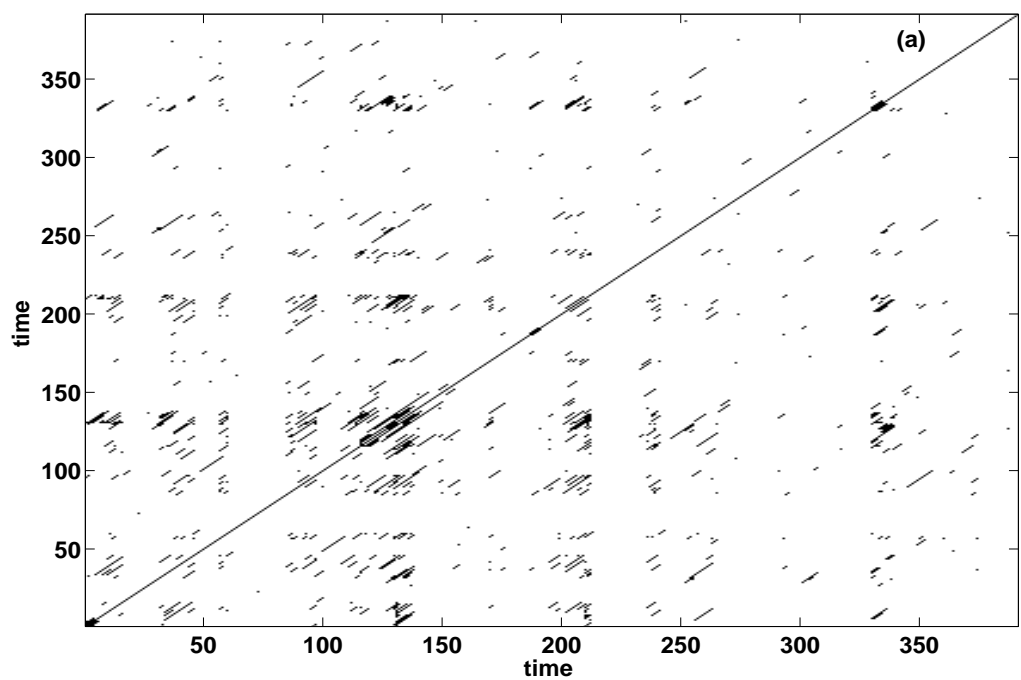}}
\resizebox{16cm}{10cm}
                {\includegraphics{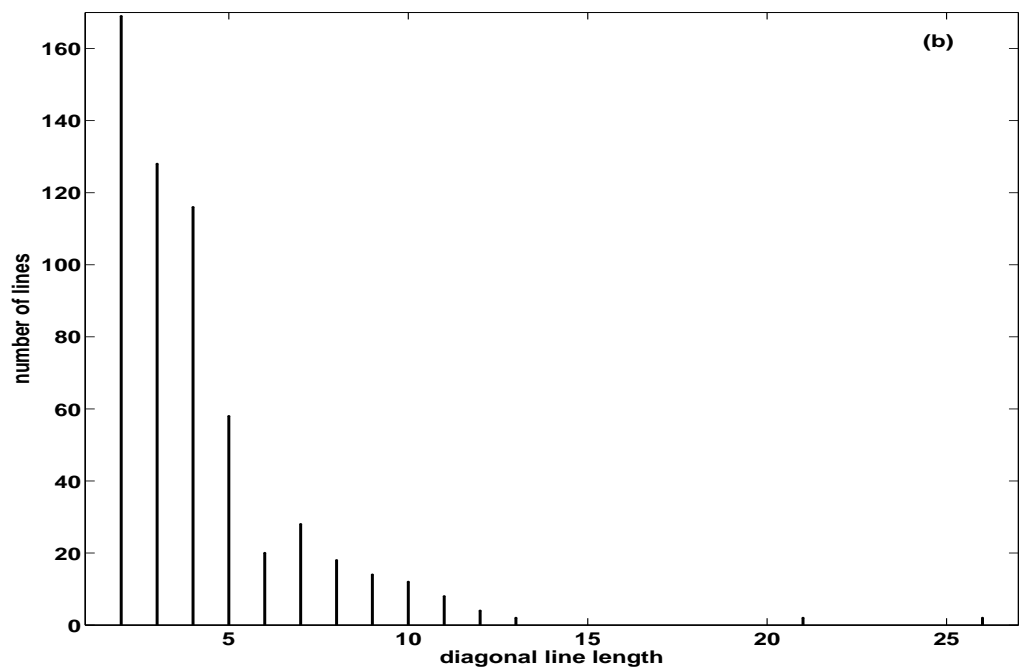}}
\caption{Recurrence plot (a) for the signal presented in Fig.2b and diagonal line distribution at (b). Time delay is equal to 1, embedding dimension 10. The values of $\gamma$ and $T$ are the same as in Fig.2}
\end{figure}

\newpage
\begin{figure}
\resizebox{16cm}{12cm}
                {\includegraphics{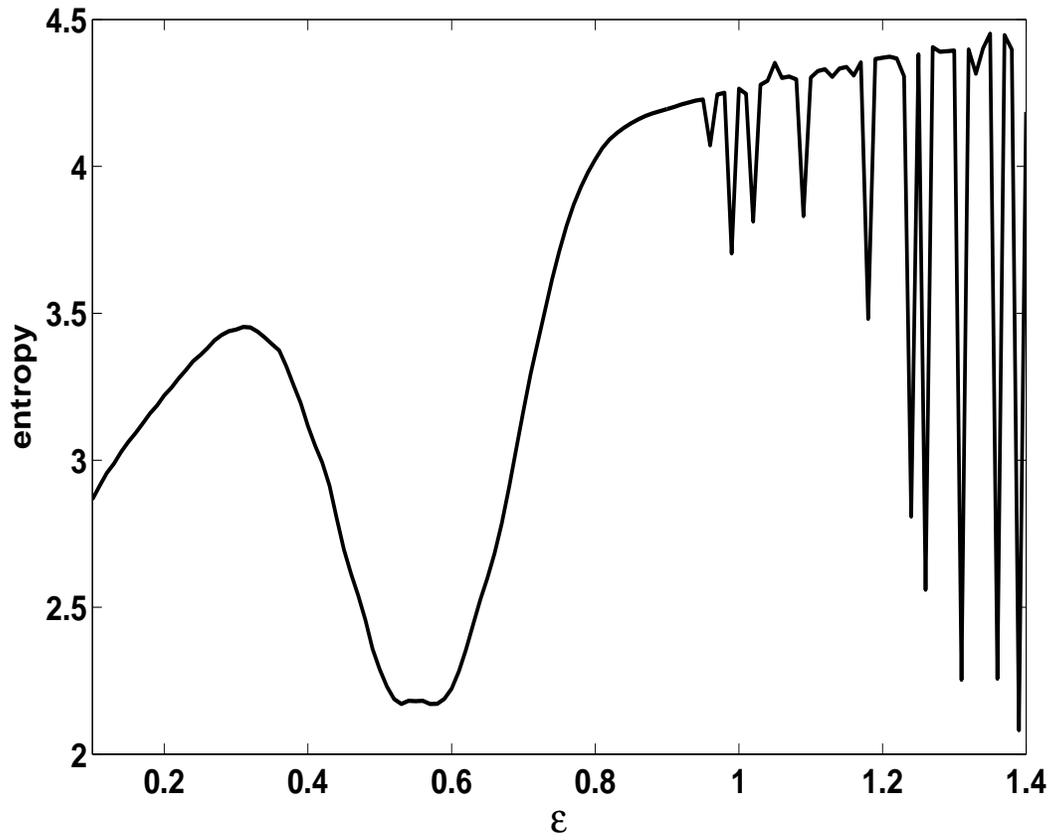}}
\caption{Entropy of changes for $\delta_n(\rho)$ versus excitation strength $\epsilon$. We assume $T=\pi$ and  $\gamma=0.1$.} 
\end{figure}

\end{document}